\documentclass[amsmath,amssymb,aps,floatfix,pre,showpacs,superscriptaddress,twocolumn]{revtex4}
\usepackage{graphicx}
\begin{document}

\title{Monte Carlo simulation of uncoupled continuous-time random walks
yielding a stochastic solution of the space-time fractional diffusion equation}

\author{Daniel Fulger}
\email{fulger@staff.uni-marburg.de}
\affiliation{Department of Chemistry and WZMW, Computer Simulation Group,
Philipps-University Marburg, 35032 Marburg, Germany}
\author{Enrico Scalas}
\email{enrico.scalas@mfn.unipmn.it}\homepage{www.mfn.unipmn.it/~scalas}
\affiliation{Department of Advanced Sciences and Technology,
Laboratory on Complex Systems, Amedeo Avogadro University of East Piedmont,
Via Vincenzo Bellini 25 G, 15100 Alessandria, Italy}
\author{Guido Germano}
\email[Corresponding author.]
{guido@staff.uni-marburg.de}\homepage{www.staff.uni-marburg.de/~germano}
\affiliation{Department of Chemistry and WZMW, Computer Simulation Group,
Philipps-University Marburg, 35032 Marburg, Germany}

\begin{abstract}
We present a numerical method for the Monte Carlo simulation of uncoupled
continuous-time random walks with a L\'evy $\alpha$-stable distribution of
jumps in space and a Mittag-Leffler distribution of waiting times, and apply it
to the stochastic solution of the Cauchy problem for a partial differential
equation with fractional derivatives both in space and in time.
The one-parameter Mittag-Leffler function is the natural survival probability
leading to time-fractional diffusion equations. Transformation methods for
Mittag-Leffler random variables were found later than the well-known
transformation method by Chambers, Mallows, and Stuck for L\'evy $\alpha$-stable
random variables and so far have not received as much attention; nor have they
been used together with the latter in spite of their mathematical relationship
due to the geometric stability of the Mittag-Leffler distribution.
Combining the two methods, we obtain an accurate approximation of space- and
time-fractional diffusion processes almost as easy and fast to compute as for
standard diffusion processes.
\end{abstract}

\date{\today}


\maketitle

\section{Introduction}
Continuous-time random walks (CTRWs) and fractional diffusion equations (FDEs),
or fractional Fokker-Planck equations, have received increasing attention.
Metzler and Klafter reviewed analytical and numerical methods to solve
fractional equations of diffusive type \cite{Metzler2000}. In
Refs.~\onlinecite{Sokolov2001,Zaslavsky2002,Barkai2002,Meerschaert2002,
Metzler2004,Flomenbom2005,Scalas2006,Zhang2006,Langlands2006}, applications and
enhancements of these techniques were presented. The relevance of fractional
calculus in the phenomenological description of anomalous diffusion has been
discussed within applications of statistical mechanics in physics, chemistry
and biology \cite{Bouchaud1990,Ott1990,Havlin2000,Castillo2005,Sokolov2006,
Dubbeldam2007a,Dubbeldam2007b} as well as finance \cite{Scalas2000a,Scalas2000b,
Mainardi2000,Cartea2007a,Cartea2007b}; even human travel and the spreading of
epidemics were modeled with fractional diffusion \cite{Brockmann2006}.
A direct Monte Carlo approach to fractional Fokker-Planck dynamics through
the underlying CTRW requires random numbers drawn from the Mittag-Leffler
distribution. Since sampling the latter was considered troublesome, different
schemes to avoid it were proposed. One possibility consists in replacing it
with the Pareto distribution---i.e., its asymptotic power-law approximation for
$t \to \infty$ \cite{Heinsalu2006}; however, as the authors point out, this is
limited to long times and an index $\beta$ not close to 1. A more general
alternative is based on subordination \cite{Gorenflo2007,Magdziarz2007a,
Magdziarz2007b}. 
Here we present a straightforward Monte Carlo method for the efficient
simulation of uncoupled CTRWs using an inversion formula for the Mittag-Leffler
distribution and apply it to compute approximate solutions of the Cauchy
problem for a generalized diffusion equation that has fractional space and time
derivatives.

\section{Theory}
\subsection{Continuous-time random walks}
\begin{figure}[h]
\includegraphics*[angle=-90,width=0.48\textwidth]{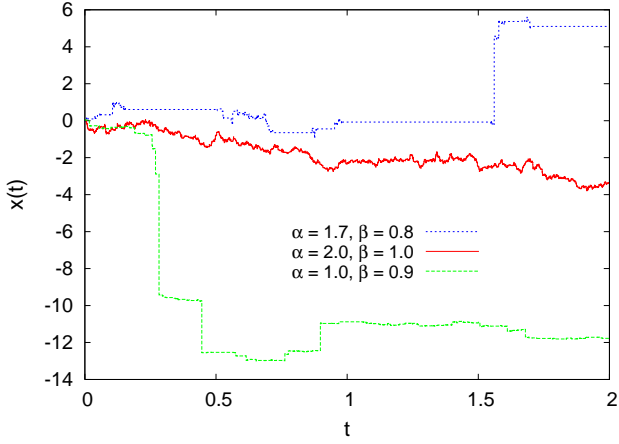}
\caption{\label{fig:walks} (Color online)
Sample paths of CTRWs with scale parameters $\gamma_t = 0.001,\ \gamma_x =
\gamma_t^{\beta/\alpha}$, and different choices of $\alpha$ and $\beta$. With
smaller $\alpha$ the jumps become larger; with smaller $\beta$ the waiting
times become longer.}
\end{figure}

A CTRW \cite{Montroll1965} is a pure jump process; it consists of a sequence of
independent identically distributed (i.i.d.) random jumps (events) $\xi_i$
separated by i.i.d.\ random waiting times $\tau_i$,
\begin{equation}
\label{eq:tn}
t_n = \sum_{i=1}^n \tau_i\, , \quad \tau_i \in \mathbb{R}_+,
\end{equation}
so that the position at time $t \in [t_n,t_{n+1})$ is given by
\begin{equation}
\label{eq:xoft}
x(t) = \sum_{i=1}^{n}\xi_i, \quad \xi_i \in \mathbb{R} .
\end{equation}
A realization of the process is a piecewise constant function resulting from a
sequence of up or down steps with different height and depth; see
Fig.~\ref{fig:walks}. Jumps are assumed to happen instantaneously or at least
in negligible time. In general, jumps and waiting times depend on each other
and they can be described by a joint probability density $\varphi (\xi, \tau)$.
The latter appears in the integral equation giving the probability density
$p(x,t)$ for the process being in position $x$ at time $t$, conditioned on the
fact that it was in position $x=0$ at time $t=0$:
\begin{equation}
\label{eq:master}
p(x,t) = \delta(x)\, \Psi(t) + \int_{-\infty}^{+\infty} \!\! d\xi \int_0^t \!\!
d\tau \, \varphi(\xi,\tau)\, p(x-\xi,t-\tau) .
\end{equation}
Here the initial condition $x(0)=0$ is contained implicitly in the first term
$\delta(x)\Psi(t)$, where we find the complementary cumulative distribution
function (survival function)
\begin{equation}
\label{eq:survival}
\Psi(t) = 1 - \int_{-\infty}^{+\infty}d\xi \int_0^t d\tau \, \varphi(\xi,\tau).
\end{equation}
Recently, one of the authors of this paper presented an analytical solution of
the integral equation in the uncoupled case---i.e., when $\varphi(\xi,\tau) =
\lambda(\xi)\psi(\tau)$, where $\lambda(\xi)$ is the jump marginal density and
$\psi(\tau)$ is the waiting time marginal density \cite{Scalas2004a}.

\subsection{Fractional diffusion equation}
\begin{figure}[h]
\includegraphics*[angle=-90,width=0.48\textwidth]{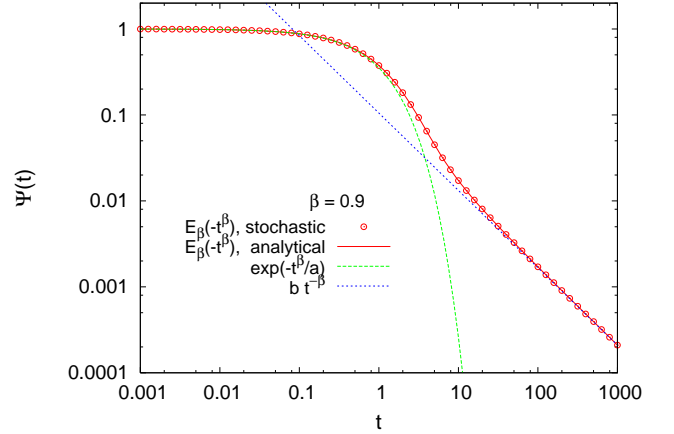}
\caption{\label{fig:ml} (Color online)
The Mittag-Leffler complementary cumulative distribution function sampled from
Eq.~(\ref{eq:Kozubowski}) (circles) and computed analytically (solid line)
\cite{Podlubny2005}, as well as its approximations for $t \to 0$ (Weibull
function, long dashes) and $t \to \infty$ (power law, short dashes).}
\end{figure}

The well-known standard diffusion equation
\begin{eqnarray}
\label{eq:standiff}
\frac{\partial}{\partial t} u(x,t) & = & D
\frac{\partial^2}{\partial x^2} u(x,t), \\ \nonumber
u(x,0^+) & = & \delta (x), \quad x \in \mathbb{R}, \quad t \in \mathbb{R}_+,
\end{eqnarray}
can be generalized to the space-time fractional diffusion equation
\begin{eqnarray}
\label{eq:fracdiff}
\frac{\partial^\beta}{\partial t^\beta} u(x,t) & = & D
\frac{\partial^\alpha}{\partial |x|^\alpha} u(x,t) \\ \nonumber
u(x,0^+) & = & \delta (x), \quad x \in \mathbb{R}, \quad t \in \mathbb{R}_+,
\end{eqnarray}
where, for $0 < \alpha \leq 2$, $\partial^\alpha / \partial |x|^\alpha$ denotes
the symmetric Riesz-Feller operator of symbol $-|\kappa^\alpha|$ and, for $0 <
\beta \leq 1$, $\partial^\beta / \partial t^\beta$ is the Caputo derivative
\cite{Caputo1971,Saichev1997,Scalas2004a}. Without loss of generality, we
assume $D = 1$; a different value would just mean a scale transformation of
space and/or time units. $u(x,t) \ge 0$ is the Green function of the FDE,
\begin{equation}
\label{eq:greenfunction}
u(x,t) = t^{-\beta/\alpha}\,W(x/t^{\beta/\alpha};\,\alpha,\beta),
\end{equation}
with the scaling function
\begin{equation}
\label{eq:scaling}
W(\xi;\,\alpha,\beta)
= \mathcal{F}^{-1}_\kappa\left[E_\beta(-|\kappa|^\alpha)\right](\xi).
\end{equation}
$E_\beta(z)$ is the one-parameter Mittag-Leffler function \cite{Hilfer2006},
\begin{equation}
\label{eq:MLpowerseries}
E_\beta(z) = \sum_{n=0}^\infty \frac{z^n}{\Gamma(\beta n+1)}, \quad
z \in \mathbb{C},
\end{equation}
with
\begin{equation}
\label{eq:MLLaplace}
E_\beta(-t^\beta)
= \mathcal{L}^{-1}_s\left[\frac{s^{\beta-1}}{1+s^\beta}\right](t), \quad
t \in \mathbb{R}_+.
\end{equation}
$\mathcal{F}$ and $\mathcal{L}$ denote the Fourier and Laplace transforms:
\begin{gather}
\widehat{f}(\kappa) = \mathcal{F}_x[f(x)](\kappa) =
\int_{-\infty}^{+\infty}f(x) e^{i \kappa x}\,dx,\\
\widetilde{f}(s) = \mathcal{L}_t[f(t)](s) = \int_0^\infty f(t) e^{-st}\,dt,
\quad s \in \mathbb{C}.
\end{gather}
For $t \in \mathbb{R}$ and $\beta = 1$, the Mittag-Leffler function with
argument $-t^\beta$ reduces to a standard exponential decay $e^{-t}$;
when $0 < \beta < 1$, the Mittag-Leffler function is approximated for small
values of $t$ by a stretched exponential decay (Weibull function)
$\exp(-t^\beta/a)$, where $a = \Gamma(\beta+1)$, and for large values of $t$ by
a power law $b t^{-\beta}$, where $b =\Gamma(\beta)\sin(\beta\pi)/\pi$; see
Fig.~\ref{fig:ml}. The Mittag-Leffler distribution is an important example of
fat-tailed waiting times; it arises as the natural survival probability
leading to time-fractional diffusion equations. There is increasing evidence
for physical phenomena \cite{Shlesinger1993,Ward1998,Mega2003} and human
activities \cite{Raberto2002,Scalas2004b,Barabasi2005} that do not follow
either exponential or, equivalently, Poissonian statistics.

Equations~(\ref{eq:greenfunction}) and (\ref{eq:scaling}) can be obtained by
Fourier-Laplace transformation of the FDE, recalling the definition of the
fractional derivatives used in Eq.~(\ref{eq:fracdiff}).

The space-fractional derivative of order $\alpha \in (0,2]$ is defined
according to Riesz \cite{Samko1993}:
\begin{equation}
\label{eq:Riesz}
\frac{d^\alpha}{d|x|^\alpha} f(x)
= \mathcal{F}^{-1}_\kappa\left[-|\kappa|^\alpha \widehat{f}(\kappa)\right](x).
\end{equation}
For $\alpha = 2$ this reduces to the usual second order derivative.
For $\alpha < 2$ the following equation holds:
\begin{equation}
\label{eq:sfe}
\frac{d^\alpha f(x)}{d|x|^\alpha}\!=\!\frac{\Gamma(\alpha\!+\!1)}{\pi}\sin
\frac{\alpha\pi}{2}\!\!\int_0^\infty\!\frac{f(x\!+\!\xi)\!-\!2f(x)\!+\!f(x\!-\!
\xi)}{\xi^{\alpha+1}}d\xi.
\end{equation}

The time-fractional derivative of order $\beta\in(0,1]$ is defined according to
Caputo \cite{Gorenflo1997,Podlubny1999}:
\begin{equation}
\label{eq:Caputo}
\frac{d^\beta}{dt^\beta}f(t)
= \mathcal{L}^{-1}_s\left[s^\beta\widetilde{f}(s)-s^{\beta-1}f(0^+)\right](t).
\end{equation}
For $\beta = 1$ this reduces to the usual first order derivative.
For $\beta < 1$ the following equation holds:
\begin{equation}
\label{eq:tfe}
\frac{d^\beta f(t)}{dt^\beta} = \frac{1}{\Gamma(1-\beta)} \left[ \frac{d}{dt}
\int_0^t \frac{f(\tau)}{(t-\tau)^\beta}\,d\tau-\frac{f(0^+)}{t^\beta} \right],
\end{equation}
where $f(0^+)$ is the initial condition.
For $\alpha=2$ and $\beta=1$, the standard diffusion equation,
Eq.~(\ref{eq:standiff}), is recovered.

It is inevitable to solve numerically a FDE in the most general case, also
known as fractional Fokker-Planck equation, which may include space- and
time-dependent diffusion and drift terms. Possible approaches are the direct
calculation of the integrals in Eqs.~(\ref{eq:sfe}) and (\ref{eq:tfe})
\cite{Ford2006}, finite-difference methods \cite{Meerschaert2006,Tadjeran2006,
Castillo2006}, and stochastic methods \cite{Meerschaert2002,Zhang2006,
Heinsalu2006,Magdziarz2007a,Magdziarz2007b}. All of them are complicated, the
latter ones mainly because of the supposedly cumbersome generation of
Mittag-Leffler random numbers. While this problem has been often worked around
in the past, we show how to overcome it, obtaining a fast and accurate method
for the Monte Carlo solution of FDEs via uncoupled CTRWs. As a benchmark, we
focus our attention on the Cauchy problem defined in Eq.~(\ref{eq:fracdiff}),
for which an analytical solution given by Eqs.~(\ref{eq:greenfunction}) and
(\ref{eq:scaling}) is available.

\subsection{Link between continuous-time random walks and the fractional
diffusion equation}
The link between CTRWs and time-fractional diffusion was discussed rigorously
in Ref.~\onlinecite{Hilfer1995} in terms of the generalized Mittag-Leffler
function $E_{\beta,\beta}(-\tau^\beta)$.

In order to approximate the Green function in Eq.~(\ref{eq:greenfunction}), it
is sufficient to simulate CTRWs whose jumps are distributed according to the
symmetric L\'evy $\alpha$-stable probability density (which reduces to a
Gaussian for $\alpha = 2$)
\begin{equation}
\label{eq:Levy}
L_\alpha(\xi) = \mathcal{F}^{-1}_\kappa\left[
\exp\left(-|\gamma_x\kappa|^\alpha\right)\right](\xi)
\end{equation}
and whose waiting times have the probability density
\begin{equation}
\label{eq:WTdensity}
\psi_\beta(\tau) = -\frac{d}{d\tau}E_\beta\left(-(\tau/\gamma_t)^\beta\right),
\end{equation}
where $E_\beta(z)$ is the one-parameter Mittag-Leffler function given by
Eq.~(\ref{eq:MLpowerseries}).
Then a weak-limit approximation of the Green function is obtained by rescaling
waiting times by a constant $\gamma_t$ and jumps by a constant $\gamma_x =
\gamma_t^{\beta/\alpha}$, letting $\gamma_t$ (and as a consequence $\gamma_x$)
vanish, and plotting the histogram for the probability density
$p_{\gamma_x,\gamma_t}(x,t;\,\alpha,\beta)$ of finding position $x$ at time $t$
for the rescaled process. This probability density weakly converges to the
Green function $u(x,t;\,\alpha,\beta)$. Weak convergence means that for $x = 0$
a singularity is always present in $p_{\gamma_x,\gamma_t}(x,t;\,\alpha,\beta)$
at $x = 0$ for any finite value of $\gamma_t$ and $\gamma_x$. This singularity
is the term $\delta(x) \Psi(t)$ in Eq.~(\ref{eq:master}) with $\Psi(t) =
E_\beta (-t^\beta)$. In the case $\alpha = 2$ and $\beta = 1$ the CTRWs are
normal compound Poisson processes (NCPPs) and, in the diffusive limit, one
recovers the Green function for the standard diffusion equation,
Eq.~(\ref{eq:standiff})---i.e., the Wiener process. This procedure is justified
in Refs.~\onlinecite{Scalas2006} and \onlinecite{Scalas2004a}. In the latter
reference, one can also find a theoretical justification for the Monte Carlo
procedure where waiting times are generated according to a power-law
distribution; a more complete treatment has been given in
Ref.~\onlinecite{Gorenflo2007}.

\section{Transformation formulas for non uniform random numbers}
The usual methods for generating random numbers with a specific probability
density are transformation, also called inversion because it requires the
inverse cumulative distribution function \cite{Feller1957}, and von Neumann
rejection \cite{vonNeumann1951}. While the latter is more general, the former
is usually faster when it is available.

\subsection{Symmetric L\'evy $\alpha$-stable probability distribution}
The symmetric L\'evy $\alpha$-stable probability density $L_\alpha(\xi)$ for
the jumps, Eq.~(\ref{eq:Levy}), can be calculated by series expansion, which we
do not report here, by direct integration \cite{Nolan1997,Nolan1999} or by
numerical Fourier transform \cite{Mittnik1999}. These methods produce a
pointwise representation of the density on a finite interval that can be used
for rejection, most efficiently with a lookup table and interpolation. More
convenient is the following transformation method by Chambers, Mallows, and
Stuck \cite{Chambers1976}:
\begin{equation}
\label{eq:Chambers}
\xi_\alpha = \gamma_x\left(\frac{-\log u \cos\phi}{\cos((1-\alpha)\phi)}
\right)^{1-1/\alpha} \frac{\sin(\alpha\phi)}{\cos\phi},
\end{equation}
where $\phi = \pi(v-1/2)$,\ $u,v \in (0,1)$ are independent uniform random
numbers, $\gamma_x$ is the scale parameter, and $\xi_\alpha$ is a symmetric
L\'evy $\alpha$-stable random number. For $\alpha=2$, Eq.~(\ref{eq:Chambers})
reduces to $\xi_2 = 2\gamma_x\sqrt{-\log u}\sin\phi$, i.e.\ the Box-Muller
method for Gaussian deviates. The other two notable limit cases are the Cauchy
distribution, with $\alpha=1$ and $\xi_1 = \gamma_x \tan\phi$, and the L\'evy
distribution, with $\alpha=1/2$ and $\xi_{1/2} = - \gamma_x \tan\phi / (2\log u
\cos\phi)$.


\subsection{One-parameter Mittag-Leffler probability distribution}
The probability density $\psi_\beta(\tau)$ for the waiting times,
Eq.~(\ref{eq:WTdensity}), can be computed as a power series from the definition
of the one-parameter Mittag-Leffler function, Eq.~(\ref{eq:MLpowerseries}),
leading to a pointwise representation on a finite interval; random numbers can
then be produced by rejection, again with a lookup table and interpolation.
Though CTRW sample paths with a Mittag-Leffler waiting time distribution have
appeared in the literature \cite{Gorenflo2004,Gorenflo2007,Magdziarz2007a,
Magdziarz2007b}, so far it has not been recognized in this context that
inversion formulas analogous to Eq.~(\ref{eq:Chambers}) are available
\cite{Devroye1996,Pakes1998,Kozubowski1998,Kozubowski1999,Kozubowski2000,
Kozubowski2001,Jayakumar2003,Germano2006}. The most convenient expression is
due to Kozubowski and Rachev \cite{Kozubowski1999}:
\begin{equation}
\label{eq:Kozubowski}
\tau_\beta = -\gamma_t\log u
\left(\frac{\sin(\beta\pi)}{\tan(\beta\pi v)}-\cos(\beta\pi)\right)^{1/\beta},
\end{equation}
where $u, v \in (0,1)$ are independent uniform random numbers, $\gamma_t$ is
the scale parameter, and $\tau_\beta$ is a Mittag-Leffler random number. For
$\beta=1$, Eq.~(\ref{eq:Kozubowski}) reduces to the inversion formula for the
exponential distribution: $\tau_1 = -\gamma_t\log u$.
Equation~(\ref{eq:Kozubowski}) and equivalent forms stem from mixture
representations of a Mittag-Leffler random variable through an exponential and
a stable random variable. The oldest representation is \cite{Devroye1996,
Jayakumar2003}
\begin{equation}
\label{eq:MLrepresentation1}
\tau_\beta = \tau_1^{1/\beta} \, \xi_{\beta,1},
\end{equation}
where $\xi_{\beta,1}$ is a skew L\'evy $\alpha$-stable random number
independent of $\tau_1$, with index $\alpha = \beta$, skewness parameter 1,
and scale factor $\gamma_x = 1/8$. A more recent representation is
\cite{Pakes1998,Kozubowski1998}
\begin{equation}
\label{eq:MLrepresentation2}
\tau_\beta = \tau_1 \, \xi_{1+}^{\pm 1/\beta},
\end{equation}
where $\xi_{1+}$ is a positive random number distributed according to a Cauchy
distribution $L_{1+}(\xi)$ with scale parameter $\gamma_x = \sin(\beta\pi)$,
location parameter $\delta = -\cos(\beta\pi)$, and normalization on
$\mathbb{R}_+$: $L_{1+}(\xi) = L_1(\xi)/\beta$ for $\xi > 0$.

The connection of Mittag-Leffler to stable random variables can be obtained in
the framework of the theory of geometric stable distributions. A random variable
$\xi$ is stable if and only if, for all $n \in \mathbb{N}$ i.i.d.\ copies of
it, $\xi_1, \ldots, \xi_n$, there exist constants $a_n \in \mathbb{R}_+$ and
$b_n \in \mathbb{R}$ such that the scaled and shifted sum $a_n (\xi_1 + \cdots
+ \xi_n) + b_n $ has the same distribution as $\xi$. A Mittag-Leffler random
variable is not stable, but it is geometric stable \cite{Kotz2001}; i.e., it is
the weak limit for $p \to 0$ of the appropriately scaled and shifted geometric
random sum $a(p)[\tau_1 + \cdots + \tau_{\nu(p)}] + b(p)$ of suitable i.i.d.\
random variables $\tau_i$, where $\nu(p)$ is a geometric random variable
indepedent of each $\tau_i$, with mean $1/p,\ p \in (0,1),$ and a geometric
probability distribution
\begin{equation}
\label{eq:geometricprob}
P(\nu(p) = n) = p (1-p)^{n-1}, \quad n \in \mathbb{N}.
\end{equation}
A random variable is geometric stable if and only if its characteristic
function $\widehat{\psi}(\kappa)$ is related to the characteristic function
$\widehat{\lambda}(\kappa)$ of a stable random variable by the equation
\cite{Mittnik1991}
\begin{equation}
\label{eq:correspondence}
\widehat{\psi}(\kappa) = \frac{1}{1-\log\widehat{\lambda}(\kappa)}.
\end{equation}
With this one-to-one correspondence, a parametrization of a geometric stable
probability density $\psi(x)$ can be established from a parametrization of the
corresponding stable probability density $\lambda(x)$.
Geometric random sums of symmetric $\tau_i$ yield the class of Linnik
distributions (a generalization of the Laplace distribution
$\frac{1}{2}e^{-|t|}$), while positive $\tau_i$ yield the class of
Mittag-Leffler distributions (as already seen, a generalization of the
exponential distribution $e^{-t},\ t \geq 0$). In particular, the
Mittag-Leffler distribution can be written as a mixture of exponential
distributions \cite{Gorenflo1997,Kozubowski2001}:
\begin{equation}
\label{eq:MLmixture}
E_\beta (-t^\beta) = \int_0^\infty \exp(-\mu t) g(\mu) \, d\mu,
\end{equation}
with a weight
\begin{equation}
\label{eq:MLweight}
g(\mu) = \frac{1}{\pi} \frac{\sin(\beta\pi)}{\mu^{1+\beta}
+ 2 \cos(\beta\pi) \mu + \mu^{1-\beta}}
\end{equation}
given by $g(\mu)d\mu = L_{1+}(\mu^\beta)d\mu^\beta$, where $L_{1+}(\xi)$ is the
probability density of $\xi_{1+}$ in Eq.~(\ref{eq:MLrepresentation2})
introduced before. Equations~(\ref{eq:MLmixture}) and (\ref{eq:MLweight})
express Eq.~(\ref{eq:MLrepresentation2}) in terms of density functions.
The inverse cumulative distribution of $L_{1+}(\xi)$ yields the transformation
formula for $\xi_{1+}$ appearing as the argument of the power function in
Eq.~(\ref{eq:Kozubowski}) \cite{Kozubowski1999,Kozubowski2000}.
Alternatively, the inversion formula $\xi_1 = \gamma_x \tan \phi + \delta$
for $L_1(\xi)$, see Eq.~(\ref{eq:Chambers}), can be substituted into
Eq.~(\ref{eq:MLrepresentation2}), provided negative values of $\xi_1$ are
discarded.

An older equivalent form of Eq.~(\ref{eq:Kozubowski}) was obtained substituting
an inversion formula for $\xi_{\beta,1}$ \cite{Kanter1975} into
Eq.~(\ref{eq:MLrepresentation1}) \cite{Devroye1996,Jayakumar2003}. A similar
result can be reached using a general transformation formula for skew L\'evy
$\alpha$-stable random numbers \cite{Chambers1976}, of which
Eq.~(\ref{eq:Chambers}) is a special case with skewness parameter 0. Both ways
require three independent uniform random numbers and more transcendent
functions than Eq.~(\ref{eq:Kozubowski}), making the latter slightly more
appealing from a numerical point of view.

\section{Numerical results}
Examples of CTRWs generated according to the described procedure---i.e.,
Eqs.~(\ref{eq:tn}), (\ref{eq:xoft}), (\ref{eq:Chambers}) and
(\ref{eq:Kozubowski})---are shown in Fig.~\ref{fig:walks}. The complementary
cumulative distribution function (survival function) of random numbers obtained
through Eq.~(\ref{eq:Kozubowski}) is checked against its analytic value
\cite{Podlubny2005} and its approximations for $t \to 0$ and $t \to \infty$ in
Fig.~\ref{fig:ml}, where a log-log scale and logarithmic binning
\cite{Newman2005} is used. Timings are reported in Table~\ref{tab:cpu} and
Ref.~\onlinecite{Germano2006}.


\begin{table}[t]
\begin{ruledtabular}
\begin{tabular*}{\columnwidth}{@{\extracolsep{\fill}}rrrrr}
$\alpha$ & $\beta$ & $\gamma_t$ & $\bar{n}$ & $t_\mathrm{CPU}$/sec \\
\hline
     2.0 &     1.0 &      0.010 &  200 &     337 \\ 
     2.0 &     1.0 &      0.001 & 2000 &    3362 \\ 
     1.7 &     0.8 &      0.010 &   74 &     437 \\ 
     1.7 &     0.8 &      0.001 &  470 &    2895 \\ 
\end{tabular*}
\end{ruledtabular}
\caption{\label{tab:cpu}
Average number $\bar{n}$ of jumps per run and total CPU time $t_\mathrm{CPU}$
in seconds for $10^7$ runs with $t \in [0,2]$ on a 2.2 GHz AMD Athlon 64 X2
Dual-Core with Fedora Core 4 Linux, using the \texttt{ran1} uniform random
number generator \cite{Press2003} and the Intel C++ compiler version 9.1 with
the -O3 -static optimization options.}
\end{table}

\begin{figure}[b]
\vspace*{-7mm}
\hspace*{-6mm}\includegraphics*[angle=-90,width=0.58\textwidth]{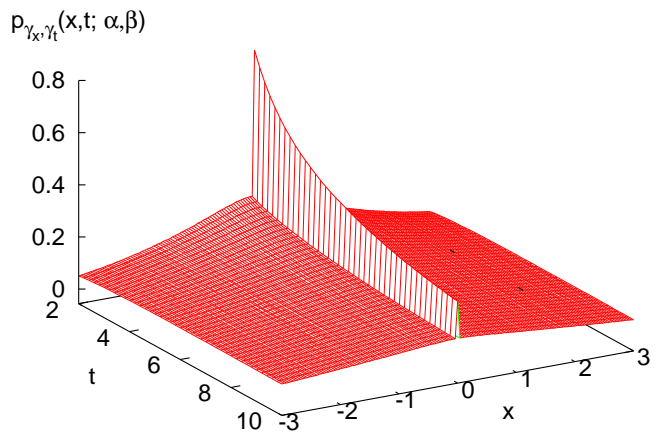}
\vspace*{-9mm}
\caption{\label{fig:decay} (Color online)
Decay of the probability density $p_{\gamma_x,\gamma_t}(x,t;\,\alpha,\beta)$
with $\alpha = 1.7,\ \beta = 0.8,\ \gamma_t = 0.1,$ and $\gamma_x =
\gamma_t^{\beta/\alpha}$. The crest at $x = 0$ is the survival function
$\Psi(t) = E_\beta\left(-(t/\gamma_t)^\beta\right) = P(0^+,t) - P(0^-,t)$,
where $P(x,t) = \int_{-\infty}^{x} p(u,t) \, du$.}
\end{figure}

\begin{figure}[t]
\includegraphics[angle=-90,width=0.48\textwidth]{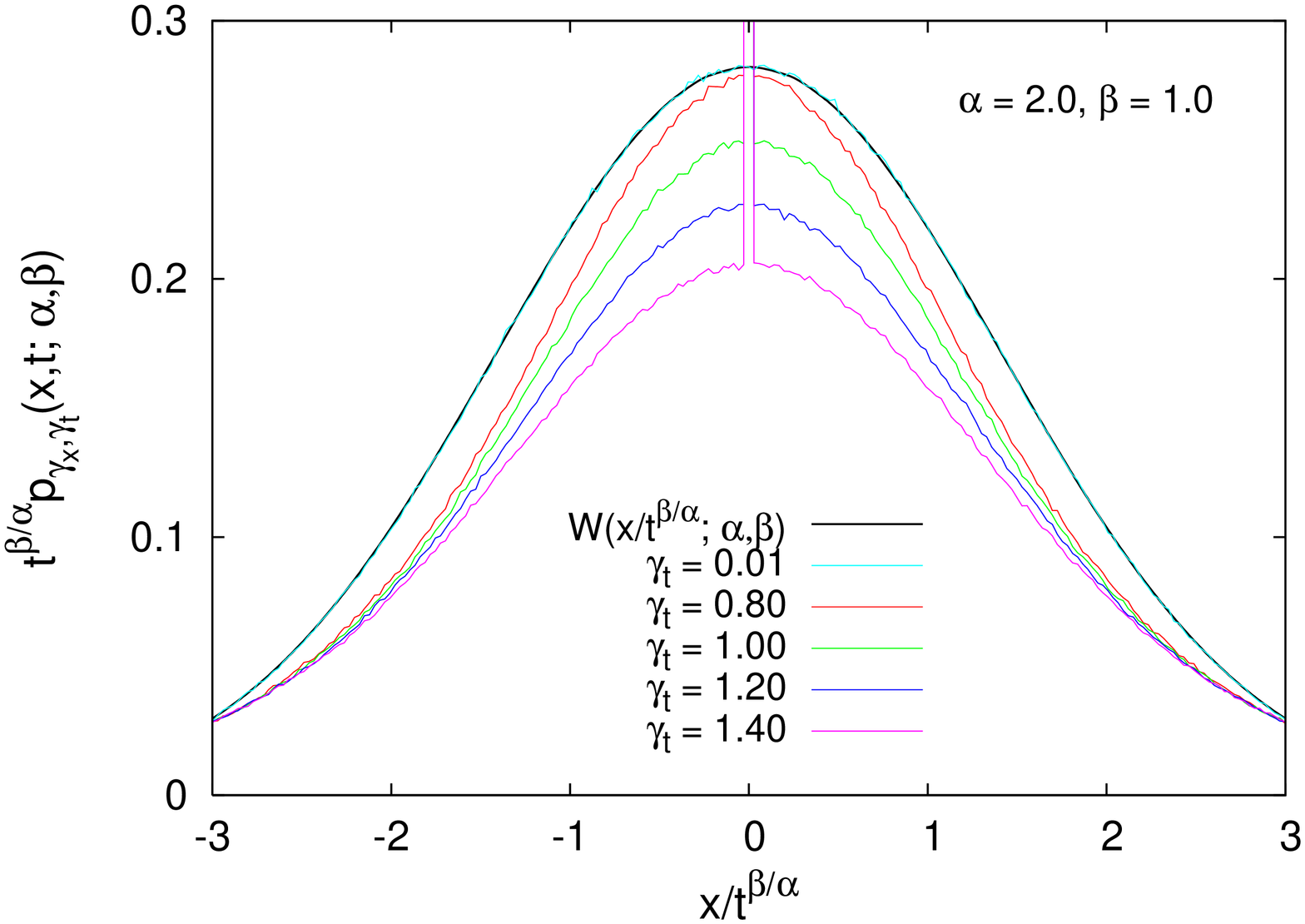}
\includegraphics[angle=-90,width=0.48\textwidth]{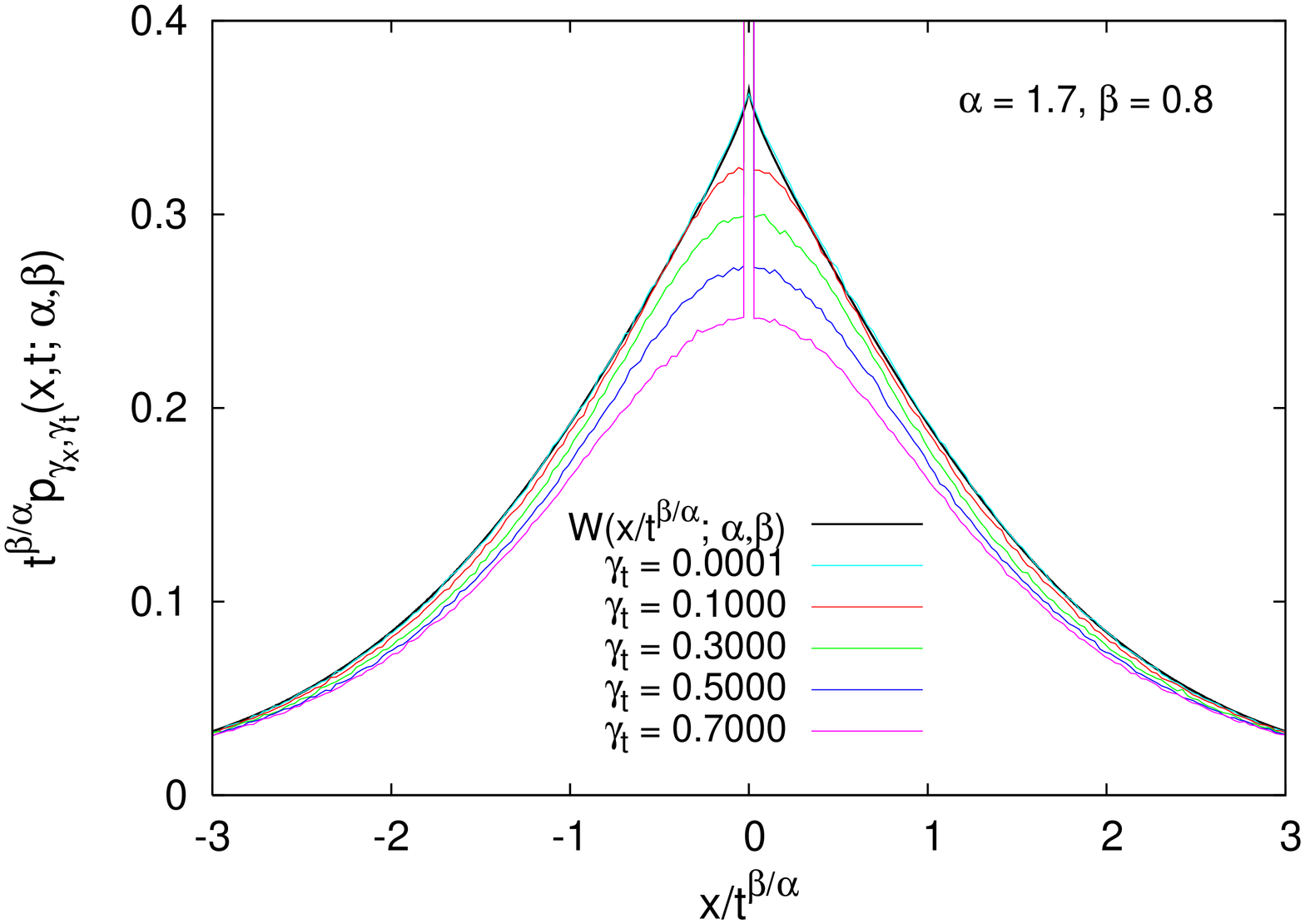}
\includegraphics[angle=-90,width=0.48\textwidth]{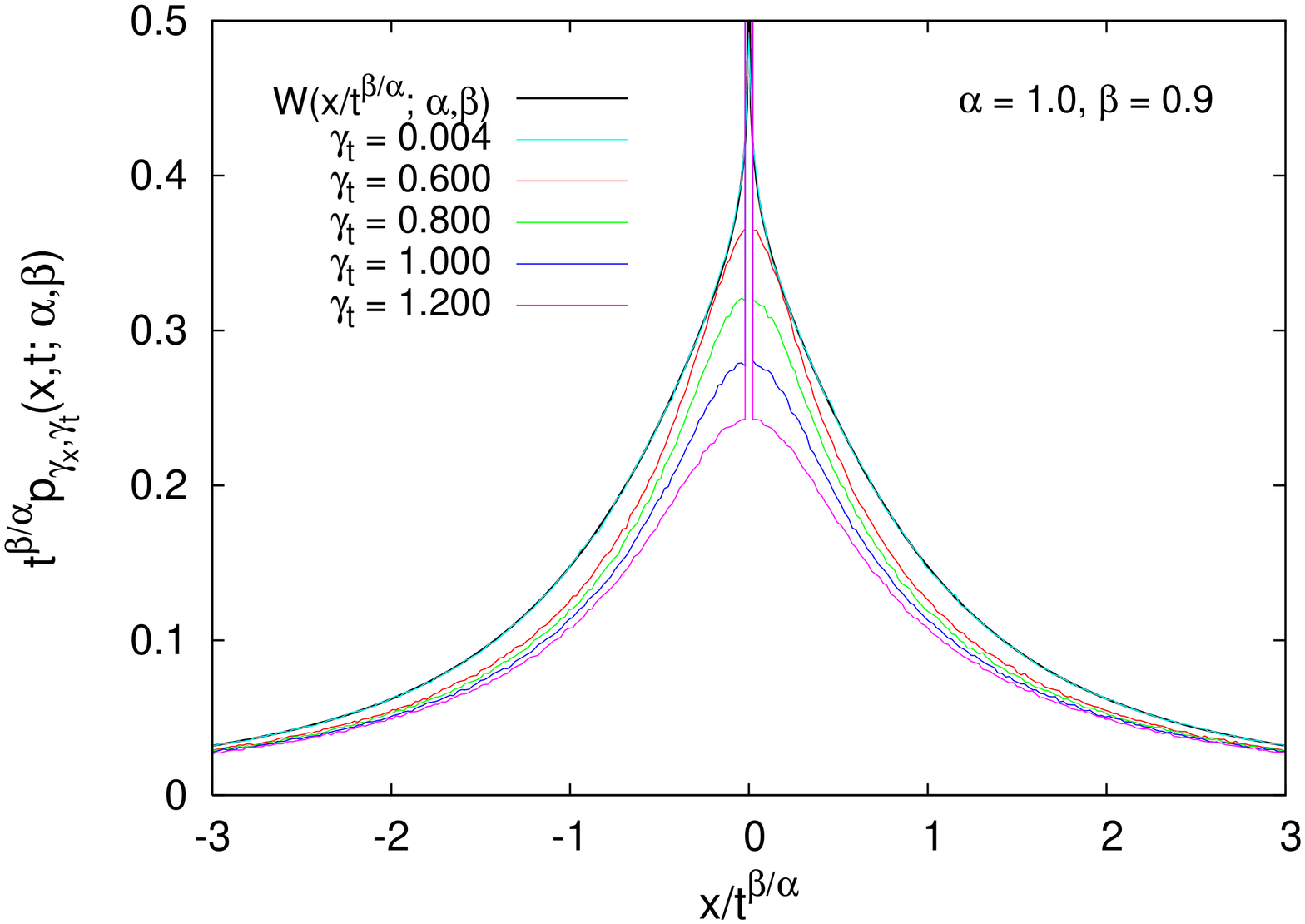}
\caption{\label{fig:convergence} (Color online)
Convergence of $t^{\beta/\alpha}p_{\gamma_x,\gamma_t}(x,t;\,\alpha,\beta)$ to
the scaling function $W(x/t^{\beta/\alpha};\,\alpha,\beta)$,
Eq.~(\ref{eq:scaling}), at $t = 2$ for selected values of $\alpha$ and
$\beta$. The curves are shown in a time-independent way as scaling plots and
appear in the same order from bottom to top as reported in the legend---i.e.,
with decreasing $\gamma_t$. The curve with the smallest $\gamma_t$ is almost
indistinguishable from its theoretical limit $W$ (solid black line). However,
in spite of the impression that may arise from the few terms and the ranges
chosen here, in general the function sequences are not monotonic. The scale
parameters $\gamma_x$ and $\gamma_t$ tend to 0 as $\gamma_x^\alpha =
\gamma_t^\beta$. The central peak decreases when the ratio $t/\gamma_t$ becomes
larger, as is evident in Fig.~\ref{fig:decay}.}
\end{figure}

\begin{figure}[t]
\includegraphics[angle=-90,width=0.48\textwidth]{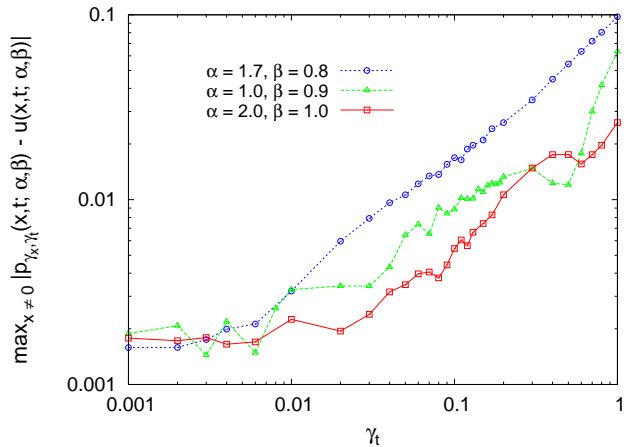}
\caption{\label{fig:norm} (Color online)
Convergence of $\max_{x\neq0}|p_{\gamma_x,\gamma_t}(x,t;\,$ $\alpha,\beta)
- u(x,t;\,\alpha,\beta)|$ for selected values of $\alpha$ and $\beta$ when
$\gamma_x,\ \gamma_t \to 0$ with $\gamma_x^\alpha = \gamma_t^\beta$.}
\end{figure}

The advantage of Eq.~(\ref{eq:Kozubowski}) is that Mittag-Leffler deviates are
generated with a simple and elegant procedure and no accuracy losses due to
truncation of the power series in Eq.~(\ref{eq:MLpowerseries}) or truncation of
the density function to a finite interval as necessary in the rejection method.
The effects of the truncation of the jump density in L\'evy flights are
analyzed in Ref.~\onlinecite{Mantegna1994}, whereas no study is available for
truncation effects on Mittag-Leffler deviates. Together with
Eq.~(\ref{eq:Chambers}), a scheme is obtained that yields sample paths for a
CTRW with a L\'evy jump marginal density and a Mittag-Leffler waiting time
marginal density at a speed comparable to that of a NCPP: Though each point for
a generic CTRW takes about 3.6 times more than for a NCPP, fewer points are
necessary (see $\bar{n}$ in Table~\ref{tab:cpu}) because the waiting times are
longer. 
The latter reference reports also that if L\'evy and Mittag-Leffler random
numbers are produced by rejection, computing the values of the probability
density functions simple-mindedly with a series expansion every time they are
needed, rather than just once at the beginning to set up a lookup table,
for L\'evy deviates the procedure takes 400 times longer than with
Eq.~(\ref{eq:Chambers}) and for Mittag-Leffler deviates it takes 5000 times
longer than with Eq.~(\ref{eq:Kozubowski}). Because of the slow convergence of
the power series in Eq.~(\ref{eq:MLpowerseries}), up to 200 terms are necessary
to achieve an acceptable accuracy, and each term is computationally expensive
because of the $\Gamma$ function. Of course these are extreme figures on the
other end of the efficiency scale meant to show how wide the latter can be;
there are smarter ways to compute both the L\'evy and Mittag-Leffler
\cite{Podlubny1999,Podlubny2005} probability densities.

Using many CTRW realizations, histograms can be built that give the evolution
of $p(x,t)$ with initial condition $p(x,0) = \delta(x)$, as displayed in
Fig.~\ref{fig:decay}. According to Eq.~(\ref{eq:master}), the initial condition
evolves as $\delta(x)\Psi(t)$; i.e., it is visible as a spike at $x=0$ that
decays as $t$ evolves. The mass of the spike is $\Psi(t) = E_\beta\left(-(t/\gamma_t)^\beta\right)$. In Fig.~\ref{fig:decay} this feature appears as a crest.
Figure~\ref{fig:convergence} shows how histograms built with CTRWs converge to
the Green function, Eq.~(\ref{eq:greenfunction}), of the FDE for decreasing
values of the scale parameters $\gamma_t$ and $\gamma_x = \gamma_t^{\beta/\alpha}$. To evaluate the scaling function in Eq.~(\ref{eq:scaling}) needed for
Eq.~(\ref{eq:greenfunction}), we used standard algorithms for
$E_\beta(-t^\beta)$ \cite{Podlubny1999,Raberto2002,Podlubny2005}, including the
fast Fourier transform. In Fig.~\ref{fig:norm} we plot $\max_{x \neq 0}
|p_{\gamma_x,\gamma_t}(x,t;\,\alpha,\beta) - u(x,t;\,\alpha,\beta)|$ as a
function of vanishing $\gamma_t$ with $\gamma_x = \gamma_t^{\beta/\alpha}$.
A rigorous analysis of convergence bounds is beyond the scope of this paper.

\section{Conclusions}
The use of Mittag-Leffler random numbers generated according to
Eq.~(\ref{eq:Kozubowski}) in combination with L\'evy random numbers generated
according to Eq.~(\ref{eq:Chambers}) is very useful in the Monte Carlo
simulation of uncoupled continuous-time random walks. In the hydrodynamic
limit, appropriately rescaled uncoupled continuous-time random walks with a
one-parameter Mittag-Leffler distribution of waiting times and a symmetric
L\'evy $\alpha$-stable distribution of jumps in space yield the Green function
of the Cauchy problem for a space-time fractional diffusion equation; we
verified this for Eq.~(\ref{eq:fracdiff}), which has an analytical solution,
Eq.~(\ref{eq:greenfunction}), as a benchmark for more difficult cases where the
diffusion and drift terms depend on space and time. We have shown that the
computational effort for a fractional diffusion process is almost as small as
for a standard diffusion process. It is true that in the same fluid limit
the Green function can be obtained too by Monte Carlo sampling of just the
asymptotic power-law tail approximations of the L\'evy and Mittag-Leffler
probability distributions, at least when the indices $\alpha$ and $\beta$ are
not close to 2 and 1, respectively. However, the neat transformation formulas
given by Eqs.~(\ref{eq:Chambers}) and (\ref{eq:Kozubowski}) are
numerically so convenient that there is no good reason for resorting to the
asymptotic approximations. Moreover, we think that, in applications,
continuous-time random walks are seen as a more fundamental model than
fractional diffusion equations, and sample paths will be generated without
taking the scale parameters $\gamma_x$ and $\gamma_t$ to the diffusive limit,
by using the approach presented in this paper.

\section*{Acknowledgments}
We thank Bj\"orn B\"ottcher and Ren\'e Schilling for help with the literature
search, Rudolf Gorenflo and Francesco Mainardi for illuminating discussions,
and Tom Kozubowski for useful comments.

\bibliography{paper}

\end{document}